# Novel Balanced Single/Dual-band Bandpass Filters Based on Circular Patch Resonator

Qiao Zhang, Chang Chen, Weidong Chen, *Member, IEEE,* Jun Ding,
and Hualiang Zhang, *Senior Member, IEEE*

*Abstract*—A new approach for designing the single-band and dual-band balanced bandpass filters (BPFs) based on the circular patch resonator is proposed in this work. By modifying the output balanced ports of the balanced BPFs to inhibit or transport $TM_{31}$ mode of the circular patch resonator under the differential-mode (DM) excitation, a single-band or dual-band balanced BPF can realized. As the odd modes could be restrained under the common-mode (CM) excitation, the balanced BPFs could achieve high CM suppression in the DM passbands. Meanwhile, several perturbing ways are introduced to improve the performance of the balanced BPFs. Specifically, perturbation vias are introduced to disturb the electric field distribution to improve the CM suppression in the DM passbands, and slots are cut on the circular patch resonators to perturb the current distribution to modify the second DM passband of the dual-band balanced BPF. Furthermore, the proposed single-band and dual-band balanced BPFs are fabricated and characterized. Good agreement can be achieved between the simulated and measured results of the fabricated balanced BPFs, which validates the feasibility of the proposed design method.

## I. INTRODUCTION

THE balanced or differential-mode (DM) microwave circuits have played an important role in microwave systems due to the advantages in terms of high robustness to undesired noise, crosstalk, and electromagnetic (EM) interference [1]. Meanwhile, the rapid development of various wireless communication systems has led to the great demand of dual-band operations. To meet these needs, various types of single-band or dual-band balanced bandpass filters (BPFs) have been proposed [2]-[24].

Manuscript submitted September 17, 2019. This work was supported by the Natural Science Foundation of China under Grant 61971392.

Q. Zhang, C. Chen, and W. Chen are with the Key Laboratory of Electromagnetic Space Information, Chinese Academy of Sciences, Hefei 230027, China, and also with the School of Information Science and Technology, University of Science and Technology of China, Hefei 230027, China (e-mail: ustczhq@mail.ustc.edu.cn; chench@ustc.edu.cn; wdchen@ustc.edu.cn).
J. Ding is with the School of Physics and Electronic Science, East China Normal University, Shanghai, 200241, China (e-mail: stevendingjun@gmail.com).
H. Zhang is with the Department of Electrical and Computer Engineering, University of Massachusetts Lowell, MA 01854 USA (e-mail: hualiang_zhang@uml.edu).

In general, transmission line-based balanced BPFs can be categorized based on the resonators, including stepped-impedance resonators (SIRs) [2]-[7], open-/short-ended coupled lines [8]-[13], slotline resonators [14] and ring resonators [4], [15]. The simple coupled SIRs are used to realize a bandstop structure in the balanced filters [2]-[4]. Moreover, the folded SIRs (FSIRs) and defected ground structures (DGS) are introduced to improve the common-mode (CM) rejection level and the DM passband selectivity [5]. In addition, more complex structures, such as combination of eight SIRs [6] and octo-section stepped-impedance ring resonator (SIRR) [7] are proposed to achieve balanced dual-band and triple-band BPFs. Furthermore, other balanced dual-band BPFs based on transmission line resonators adopt stub-loaded theory [8], multi-mode resonators [9], modified coupled-embedded resonators [10], doubly short-ended resonators [11] and asymmetrical coupled lines [12] to realize the dual-band operations. In addition, by modifying the stub-loaded-type bandstop filter branches loaded at the input and output accesses, the balanced single- or dual-band BPFs can be achieved [13]. The balanced BPFs can also be realized by employing a slotline resonator placed in the ground plane [14] or ring resonators [4], [15].

Typically, the BPFs constructed by the conventional transmission line resonators commonly suffer from the high-conductor loss and low power handing capability. To overcome these drawbacks, a class of balanced BPFs based on substrate integrated waveguide (SIW) is explored [16]-[18]. The first two modes in the SIW cavities (i.e., $TE_{101}$ and $TE_{201}$ modes) could be utilized to design a balanced dual-band BPF [16]. However, the size of the structure and the frequency tuning range are limited. Therefore, various schemes such as loading split-ring resonators (CSRRs) [17] and stacking multi-layer [18] are proposed to tackle these issues.

Besides, it is found that the patch resonators on a thin substrate not only feature the same low loss and high power handling capability as the SIW cavities, but also are much simpler and more straightforward in analysis and design [19]. Due to these advantages, several balanced BPFs based on patch resonators are proposed in recent years [20-25]. Firstly, the square patch resonator is employed to design balanced BPFs [20], [21] and a balanced-to-balanced filtering power divider [22]. Secondly, the right-angled isosceles triangular patch resonator (RAITPR) is applied to design the balanced BPFs [23], [24], which could maintain a similar field distribution to the square patch resonator but with a reduced size. Also, a balanced BPF based on two overlapped RAITPRs utilized high order modes (i.e., $TM_{200}^Z$ and $TM_{210}^Z$ modes) to achieve high CM suppression in the DM passband [25].

In a balanced BPF design, the CM rejection can be achieved by adopting a bandstop structure and employing structures where the CM signal cannot be excited or transported at all, or by introducing extra elements (e.g., the open-short-ended coupled lines, defected ground structures, and etched slots) to enhance the CM absorption or reduce the CM coupling. Meanwhile, the dual-band balanced BPFs can be realized in several different ways [1]. In the first approach, each resonator produces two controllable resonant frequencies. Most of the existing dual-band balanced BPFs are achieved in this way. In the second approach, although the resonator has only one resonant frequency, it can produce two DM passbands by introducing transmission zeros (TZs) into the middle of the bands. However, very few reported balanced BPFs so far can be flexibly changed between the single-band and dual-band operations. In [13], TZs are introduced in the middle of the bands to achieve a balanced BPF that can be functioned from single-band to dual-band. However, it suffers from problems such as large size and the passbands of the dual-band balanced BPF cannot be modified.

In this work, the circular patch resonators are adopted to achieve novel single-band and dual-band balanced BPFs. The single-band or dual-band operation of the proposed balanced BPF can be realized by inhibiting or transmitting the $TM_{31}$ mode with proper designed the output balanced ports. In addition, the position for the second DM passband of the balanced BPF can be flexibly adjusted by introducing the slots cuts on the circular patches to perturb the current distribution of the $TM_{31}$ mode under the DM excitation. Moreover, the proposed balanced BPFs feature good CM suppression in the passbands of DM. The proposed single-band and dual-band balanced BPFs are fabricated and characterized. The simulated and measured results agree very well with each other. The work is organized as follows: in section II, theoretical solutions of the circular patch resonator are derived to show the modal solutions and their field patterns; the analysis and design of the single-band/dual-band balanced BPFs are detailed in section III; section IV provides the simulated and measured results of the proposed balanced BPFs, which are also compared with the other single- and dual-band balanced BPFs reported in the literature; a short conclusion is drawn in section V.

## II. ANALYSIS OF THE CIRCULAR PATCH RESONATOR

Fig. 1 depicts the geometry of a circular patch resonator. To facilitate the analysis of the resonant modes, the Cartesian coordinate system is converted to the cylindrical coordinate system. The cylindrical components of the transverse fields can be derived from the longitudinal components ($E_z/H_z$) as [26]:

$$E_\rho = -\frac{j}{k_c^2}\left(\beta\frac{\partial E_z}{\partial \rho} + \frac{\omega\mu}{\rho}\frac{\partial H_z}{\partial \varphi}\right) \quad E_\varphi = -\frac{j}{k_c^2}\left(\frac{\beta}{\rho}\frac{\partial E_z}{\partial \varphi} - \omega\mu\frac{\partial H_z}{\partial \rho}\right)$$

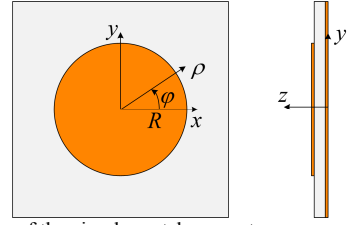

Fig. 1. Geometry of the circular patch resonator.

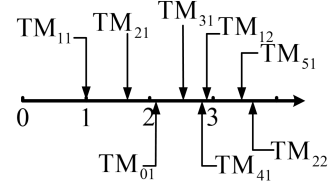

Fig. 2. Normalized resonant frequencies of the modes in the circular patch resonator.

$$H_\rho = \frac{j}{k_c^2}\left(\frac{\omega\varepsilon}{\rho}\frac{\partial E_z}{\partial \varphi} - \beta\frac{\partial H_z}{\partial \rho}\right) \quad H_\varphi = -\frac{j}{k_c^2}\left(\omega\varepsilon\frac{\partial E_z}{\partial \rho} + \frac{\beta}{\rho}\frac{\partial H_z}{\partial \varphi}\right)$$ (1)

where $\mu = \mu_0\mu_r$, $\varepsilon = \varepsilon_0\varepsilon_r$ are the permeability and permittivity of the material inside the resonator, respectively. $k_c = \sqrt{k^2 - \beta^2}$ is the cutoff wavenumber, and $k = \omega\sqrt{\mu\varepsilon}$ is the wavenumber of the dielectric material of the resonator. In this work, $e^{-j\beta z}$ propagation has been assumed.

For the TM modes in the circular patch resonator, $H_z = 0$. The wave equation can be expressed in the cylindrical coordinate system as:

$$\left(\frac{\partial^2}{\partial \rho^2} + \frac{\partial}{\rho\partial\rho} + \frac{\partial^2}{\rho^2\partial\varphi^2}\right)E_z(\rho,\varphi) + k_c^2 E_z(\rho,\varphi) = 0 \quad (2)$$

The solution of (2) for $E_z$ can be simplified as:

$$E_z(\rho,\varphi) = E_{ni}J_n(k_c\rho)\begin{cases}\cos(n\varphi)\\ \sin(n\varphi)\end{cases} \quad (n,i \in N^*) \quad (3)$$

where $J_n(x)$ is the Bessel function of the first kinds and $E_{ni}$ is an arbitrary constant.

For the circular patch resonator, the mode functions should satisfy the following boundary conditions:

$$E_\rho\big|_{\rho=R} = 0, \quad H_z\big|_{\rho=R} = 0, \quad H_\varphi\big|_{\rho=R} = 0 \quad (4)$$

where $R$ is the effective radius of the circular patch resonator. According to (1), it is obvious that $\partial E_z/\partial\rho\big|_{\rho=R} = 0$. Thus, the following condition should be satisfied:

$$J_n'(k_c\rho) = 0 \quad (5)$$

where the notation $J_n'(k_c\rho)$ refers to the derivative of $J_n$ with respect to its argument. Correspondingly, $k_c$ can be calculated as:

$$(k_c)_{ni} = \frac{v_{ni}}{R} \quad (n \in N, i \in N^*) \quad (6)$$

where $v_{ni}$ is the $i^{th}$ root of the derivative of $n^{th}$-order Bessel function [26].

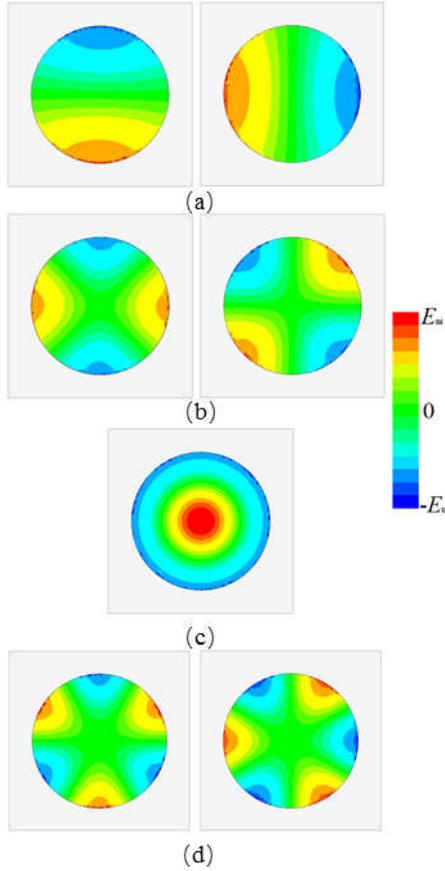

Fig. 3. $E_z$ on the circular patch resonator for different modes: (a) a pair of degenerate $TM_{11}$ modes, (b) a pair of degenerate $TM_{21}$ modes, (c) $TM_{01}$ mode (d) a pair of degenerate $TM_{31}$ modes.

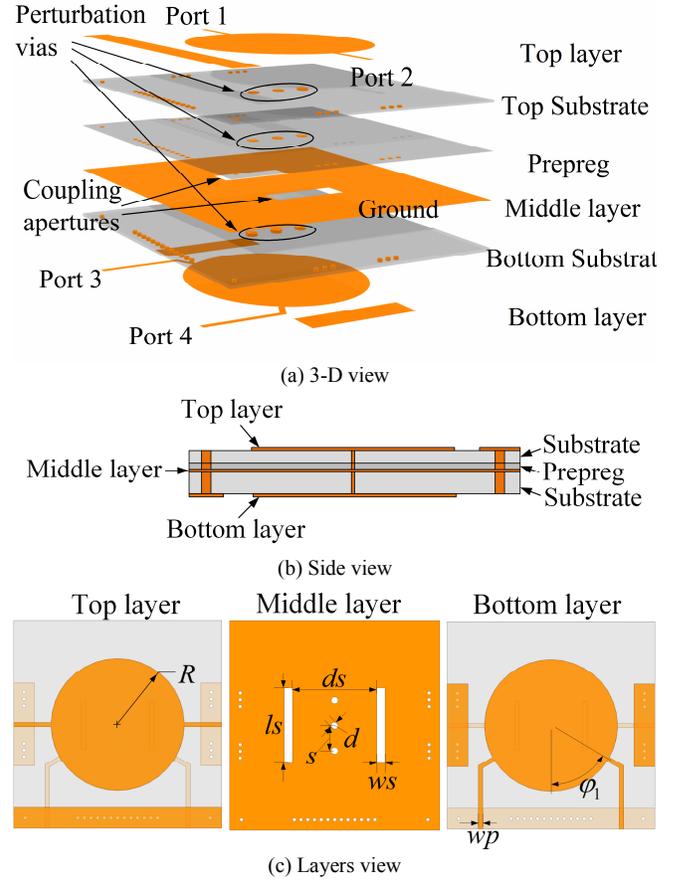

Fig. 4. Layout of the single-band balanced BPF: (a) 3-D view; (b) Side view; (c) Layers view.

The cylindrical field components for the $TM_{ni}$ mode can be calculated from (1) and (3):

$$\begin{cases} E_z = E_{ni} J_n(k_c\rho) \begin{cases} \cos(n\varphi) \\ \sin(n\varphi) \end{cases} e^{-j\beta z} \\ E_\rho = -\dfrac{j\beta E_{ni}}{k_c} J_n'(k_c\rho) \begin{cases} \cos(n\varphi) \\ \sin(n\varphi) \end{cases} e^{-j\beta z} \\ E_\varphi = -\dfrac{j\beta n E_{ni}}{k_c^2 \rho} J_n(k_c\rho) \begin{cases} -\sin(n\varphi) \\ \cos(n\varphi) \end{cases} e^{-j\beta z} \\ H_z = 0 \\ H_\rho = -\dfrac{j\omega\varepsilon n E_{ni}}{k_c^2 \rho} J_n(k_c\rho) \begin{cases} -\sin(n\varphi) \\ \cos(n\varphi) \end{cases} e^{-j\beta z} \\ H_\varphi = -\dfrac{j\omega\varepsilon E_{ni}}{k_c} J_n'(k_c\rho) \begin{cases} \cos(n\varphi) \\ \sin(n\varphi) \end{cases} e^{-j\beta z} \end{cases} \quad (7)$$

It should be noted from (7) that $TM_{ni}$ in the circular patch resonator features a pair of degenerate modes which are represented by $\sin(n\varphi)$ and $\cos(n\varphi)$. When $n=0$, $E_z = E_{0i} J_0(k_c r)$, $E_\varphi = H_\rho = 0$, there is no degenerate mode for $TM_{0i}$. From (6) and $f = \dfrac{k}{2\pi\sqrt{\mu\varepsilon}}$, the resonant frequencies of $TM_{ni}$ mode can be calculated as the following:

$$f_{ni} = \dfrac{c v_{ni}}{2\pi R \sqrt{\varepsilon_{eff}}} \quad (8)$$

where $c$ stands for the light speed in free space and $\varepsilon_{eff}$ represents the effective permittivity around the patch resonator. These resonant frequencies are normalized and plotted in Fig. 2. In this work, only the first four modes are considered. It can be seen from Fig. 2 that the frequencies of the $TM_{21}$ and $TM_{01}$ modes are between that of the $TM_{11}$ and $TM_{31}$ modes.

Fig. 3 depicts the longitudinal electric distribution of the first seven modes supported by the circular patch resonator. According to (7), there is no degeneration for $TM_{01}$ mode when $n=0$. For other situations, the pairs of degenerate modes are about $\varphi_0 = 90°/n$ center symmetry. As a result, the degenerate modes of $TM_{11}$, $TM_{21}$ and $TM_{31}$ are $90°$, $45°$ and $30°$ center symmetry, respectively, which can be clearly observed from Fig. 3.

In addition, the properties of the electric field distributions of the modes in the circular patch resonator can be mathematically expressed as:

$$\begin{cases} E_z(\rho,\varphi,z) = -E_z(\rho,\varphi+180,z) & n \text{ is odd} \\ E_z(\rho,\varphi,z) = E_z(\rho,\varphi+180,z) & n \text{ is even} \end{cases} \quad (9)$$

Thus, these modes can be categorized into odd modes (e.g.,

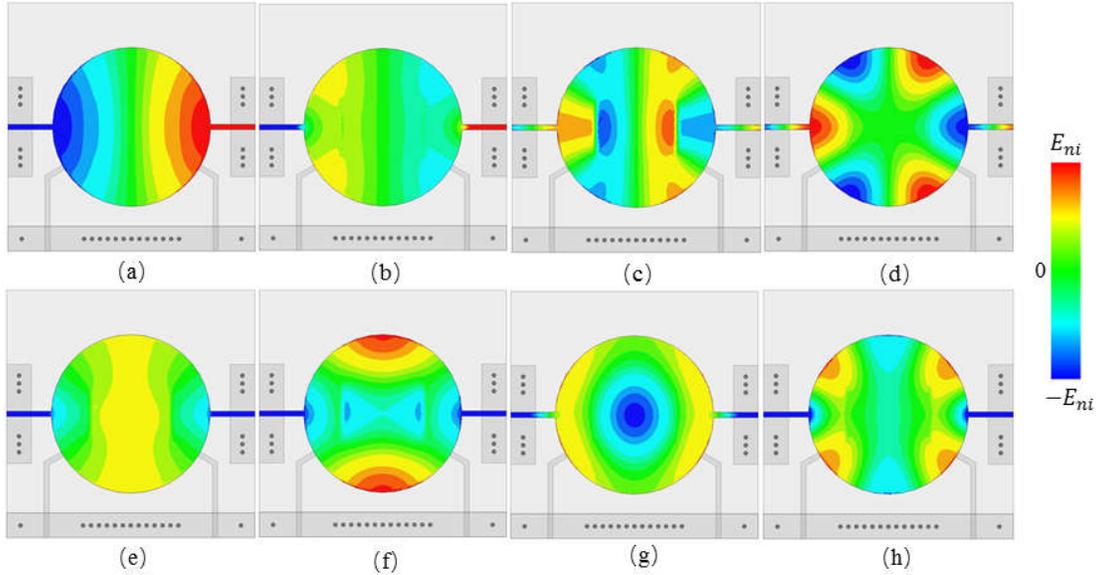

Fig. 5. $E_z$ on the upper circular patch resonator for different modes: (a) $TM_{11}$ mode (2.77 GHz), (b) $TM_{21}$ mode (4.60 GHz), (c) $TM_{01}$ mode (5.77 GHz), (d) $TM_{31}$ mode (6.33 GHz) under the DM excitation; (e) $TM_{11}$ mode, (f) $TM_{21}$ mode, (g) $TM_{01}$ mode, (h) $TM_{31}$ mode under the CM excitation.

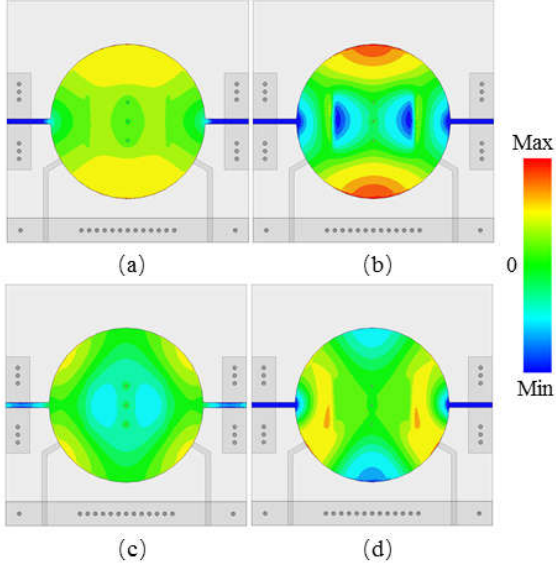

Fig. 6. $E_z$ on the upper circular patch resonator of the single-band balanced BPF for the perturbed modes under the CM excitation: (a) $TM_{11}$ mode (2.773 GHz); (b) $TM_{21}$ mode (4.735 GHz); (c) $TM_{01}$ mode (7.46 GHz); (d) $TM_{31}$ mode (6.58 GHz).

$TM_{11}$ and $TM_{31}$ modes) and even modes (e.g., $TM_{01}$ and $TM_{21}$ modes) when *n* is odd and even, respectively.

### III. ANALYSIS OF THE BALANCED BPF

As discussed in Section II, the pair of degenerate modes are center symmetry and can be classified into odd and even modes. If the positions of the feed line are chosen appropriately, the modes can be excited under the DM operation and suppressed under CM operation, which is desirable for designing the balanced BPFs. In the following, a single-band and dual-band balanced BPFs are designed by adjusting the feeding lines with additional perturbation vias and slots cuts.

#### A. Design of single-band balanced BPF

Based on the mode analysis of the circular patch resonator, a single-band balanced BPF is proposed in Fig. 4(a), which is composed of two layers of substrates with two circular patch resonators and a ground plane in the middle. As shown in Fig. 4(b), the substrates are Rogers RO4350B with a dielectric constant of 3.66 and glued by a thin layer of Rogers RO4450F prepreg with a thickness of 0.1 mm and a dielectric constant of 3.66. The thicknesses of the top and bottom substrates are 0.338mm and 0.508mm, respectively. Besides, two rectangular apertures are etched on the ground plane to realize the electric and magnetic couplings between the two patch resonators. A pair of balanced input ports (Port 1 and Port 2) are connected to the top circular patch resonator, while the pair of balanced output ports (Port 3 and Port 4) are linked to the bottom patch resonator.

Fig. 4(c) depicts that Port 1 and Port 2 are 180° center symmetry. The electric field distributions on the top resonator for different modes under the DM and CM excitations are plotted Figs. 5(a)-(d) and Figs. 5(e)-(h), respectively. It can be observed from Figs. 5(a)-(d) that the odd modes (e.g., $TM_{11}$ and $TM_{31}$ modes) could be generated under the DM excitation and the even modes (e.g., $TM_{21}$ and $TM_{01}$ modes) are restrained. In the contrary, it can be observed from Figs. 5(e)-(h) that the even modes (e.g., $TM_{21}$ and $TM_{01}$ modes) could be excited under the CM excitation and the odd modes (e.g., $TM_{11}$ and $TM_{31}$ modes) are suppressed. Therefore, under the CM excitation, a high CM suppression can be achieved in the DM passband because the odd modes are suppressed. And TZs can be introduced in the upper DM passband because the even modes are restrained under the DM excitation.

Moreover, it is noticed from Figs. 5(a) and 5(d) that $TM_{11}$ and $TM_{31}$ modes are generated simultaneously under DM excitation due to the symmetry, thus, a single-band or dual-band balanced BPF can be realized with or without suppressing the $TM_{31}$ mode transmission. For achieving a single-band balanced responses, it is required to transmit $TM_{11}$ mode only and suppress the $TM_{31}$ mode at the output balanced ports. In order to suppress $TM_{31}$ mode, the $\varphi_1$ at the bottom layer in Fig. 4(c) is chosen to be around $60°$, where the electric field distributions for $TM_{31}$ mode at the balanced output ports are almost zero.

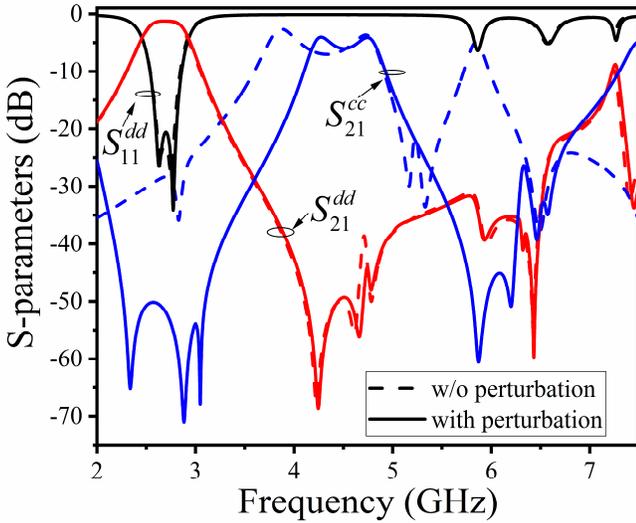

Fig. 7. The responses of the proposed single-band balanced BPF with and without perturbation.

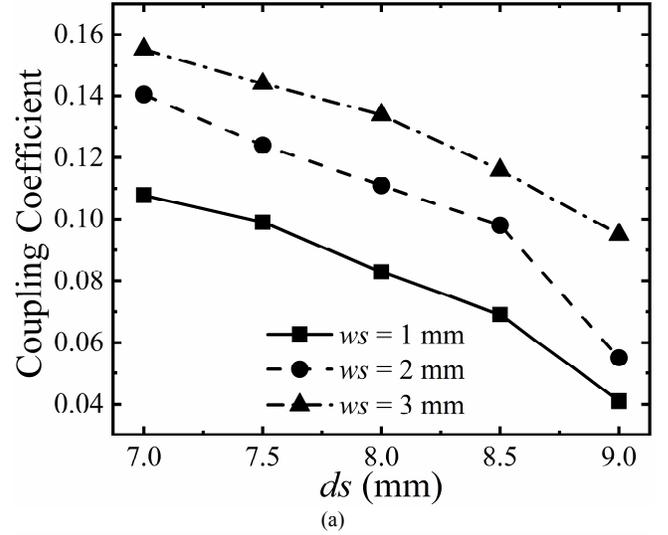

(a)

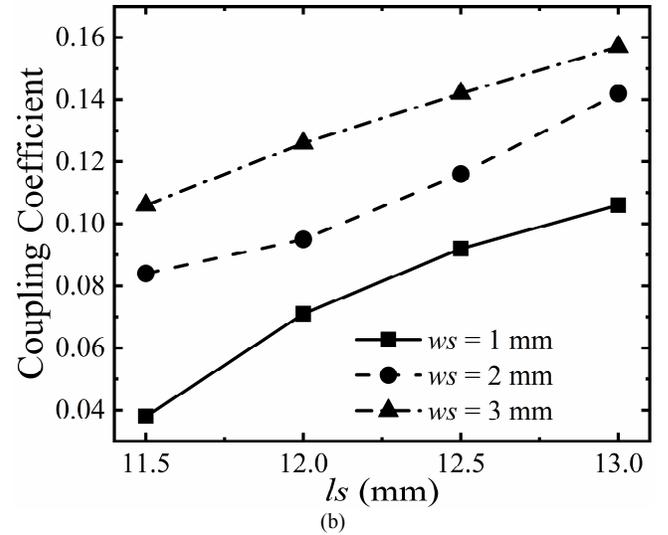

(b)

Fig. 8. The coupling coefficients of the single-band balanced BPF by varying the aperture parameters (a) $ds$ ($ls$ = 11.6 mm) and (b) $ls$ ($ds$ = 8.7 mm) with different $ws$.

Since $TM_{21}$ mode and $TM_{01}$ mode would be generated under the CM excitation, three perturbation vias shown in Fig. 4(a) are introduced to disturb the excited even modes. The perturbation vias create nulls in the electric field distribution around the vias. Therefore, the electric field distributions on the top patch resonators under the CM excitation will be distorted as shown in Fig. 6. It can be observed from Fig. 5 that the three vias have little impact to the generated odd modes under DM excitation as the three vias are located at where the electric field distributions of the odd modes are nearly zero. However, the positions of the perturbation vias are located where the electric field distributions of the even modes are strong under the CM excitation, thus, the excited modes will be distorted as shown in Fig. 6. Besides, Fig. 7 plots the simulated results of the proposed single-band balanced BPF with and without perturbation. Under the DM excitation, the performances of the balanced BPF ($S_{11}^{dd}$ and $S_{21}^{dd}$) has almost no change, but under the CM excitation, the minimal suppression in the DM passband is enhanced from 28.4 dB to 50.2 dB and the maximal CM suppression could reach up to 71 dB. Moreover, it can also be seen from Fig. 7 that resonant frequencies of the excited modes under the CM excitation will increase.

The coupling between the two circular patch resonators is realized by the two apertures etched on the ground plane. The coupling coefficient can be calculated by using the following equation [27]:

$$k = \pm \frac{f_{p2}^2 - f_{p1}^2}{f_{p2}^2 + f_{p1}^2} \qquad (10)$$

where $f_{p1}$ and $f_{p2}$ are defined as the lower and upper resonant frequencies induced by the two coupled resonators, respectively. The width and length of the apertures are $ws$ and $ls$, respectively, and the distance between the two apertures is $ds$. The simulated results by varying $ws$, $ls$ and $ds$ are plotted in Fig. 8. It can be observed from Fig. 8 that the coupling could be enhanced as $ds$ decreases or $ls$ increases or $ws$ increases.

B. *Design of dual-band balanced BPF*

As mentioned in previous section, in order to achieve dual-band balanced responses, both $TM_{11}$ and $TM_{31}$ modes are required to be transmitted to the balanced output ports. Compared to the single-band balanced BPF in Fig. 4, the dual-band balanced BPF illustrated in Fig. 9 can be achieved by changing the angle $\varphi_2$. Besides, the balanced output ports also need to be designed to suppress the generated $TM_{21}$ mode under the CM excitation to improve the CM suppression between the two DM passbands. Therefore, $\varphi_2$ in Fig. 9(c) is designed to be about $45°$, in which way that the balanced output ports are at the positions where the electric field distributions of $TM_{11}$ and $TM_{31}$ modes under the DM excitation are strong and that of $TM_{21}$ mode under the CM excitation is nearly zero. The comparisons of the single-band balanced responses and dual-band balanced responses are shown in Fig. 10. The first DM passband of the dual-band balanced BPF is located at the same position of the DM passband of the single-band balanced BPF, which is caused by the $TM_{11}$ mode. As $TM_{31}$ mode can be transported in the dual-band balanced BPF but suppressed in the single-band balanced BPF, the second DM passband is located at the resonant frequency of the $TM_{31}$ mode under the DM excitation. Moreover, the CM suppression between the two DM passbands of the dual-band balanced BPF is improved in comparison with the single-band balanced BPF.

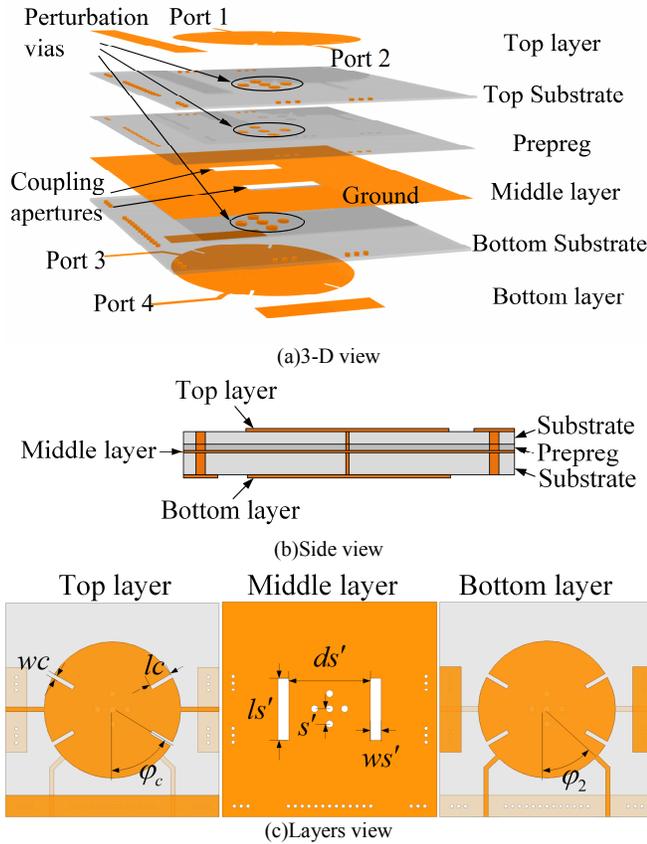

Fig. 9. Layout of the dual-band balanced BPF: (a) 3-D view; (b) Side view; (c) Layers view.

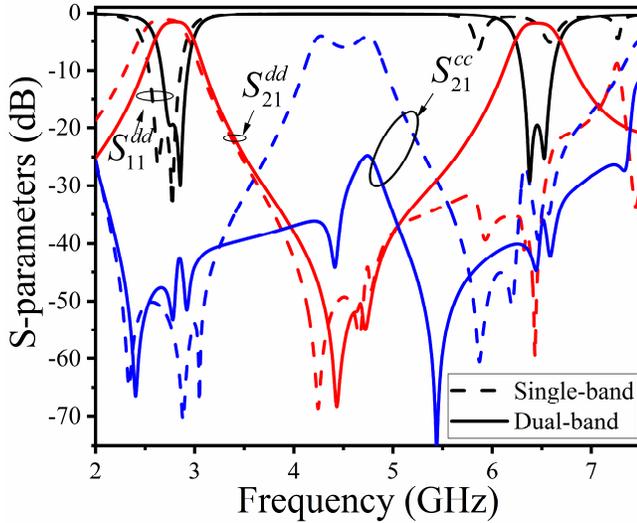

Fig. 10. Comparison between the single-band balanced and dual-band balanced responses.

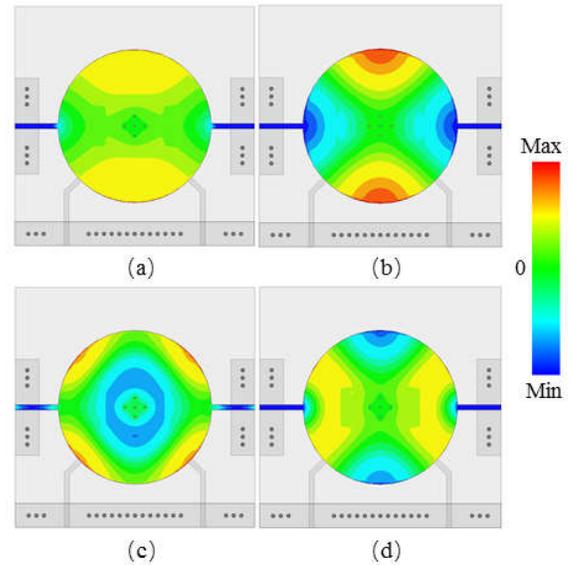

Fig. 11. $E_z$ of the perturbed modes in the upper circular patch resonator of dual-band balanced BPF under CM excitation: (a) $TM_{11}$ mode (2.855 GHz); (b) $TM_{21}$ mode (4.739 GHz); (c) $TM_{01}$ mode (7.68 GHz); (d) $TM_{31}$ mode (6.44 GHz).

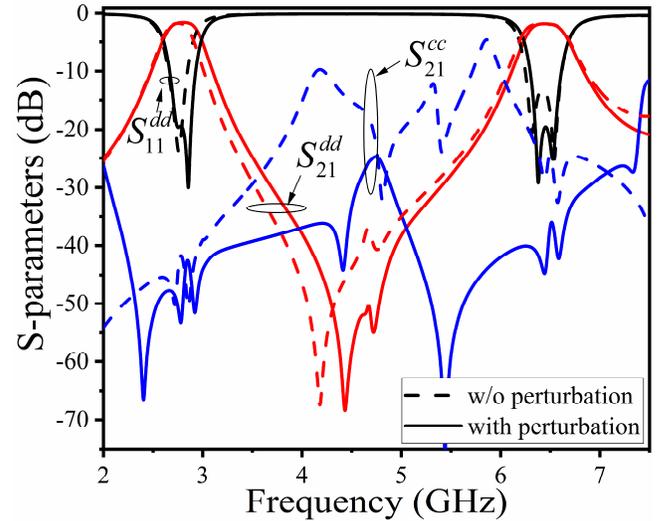

Fig. 12. The responses of the proposed dual-band balanced BPF with and without perturbation.

Furthermore, in order to improve the CM suppression around the second DM passband, as shown in Fig. 9(a), five perturbation vias which create nulls in the electric field distributions around themselves are introduced to disturb the modes under the CM excitation. The perturbation vias are located at where the electric field distributions of the odd modes under the DM excitation is nearly zero to reduce the influence on the DM responses. However, the locations of the perturbation vias are at where the electric field distributions of the even modes under the CM excitation are strong. As a result, the electric field distributions of the even modes under the CM excitation will be distorted as shown in Fig. 11, and the resonant frequencies of them will increase. Fig. 12 plots the proposed dual-band balanced BPF with and without perturbation vias. It can be seen from Fig. 12 that the minimal CM suppression in the second DM passband is increased from 20.0 dB to 36.3 dB by utilizing the cross distributed perturbation vias. In addition, the maximal CM suppression could reach up to 44.7 dB, and the minimal CM suppression between the two DM passbands is increased from 4.6 dB to 24.7 dB.

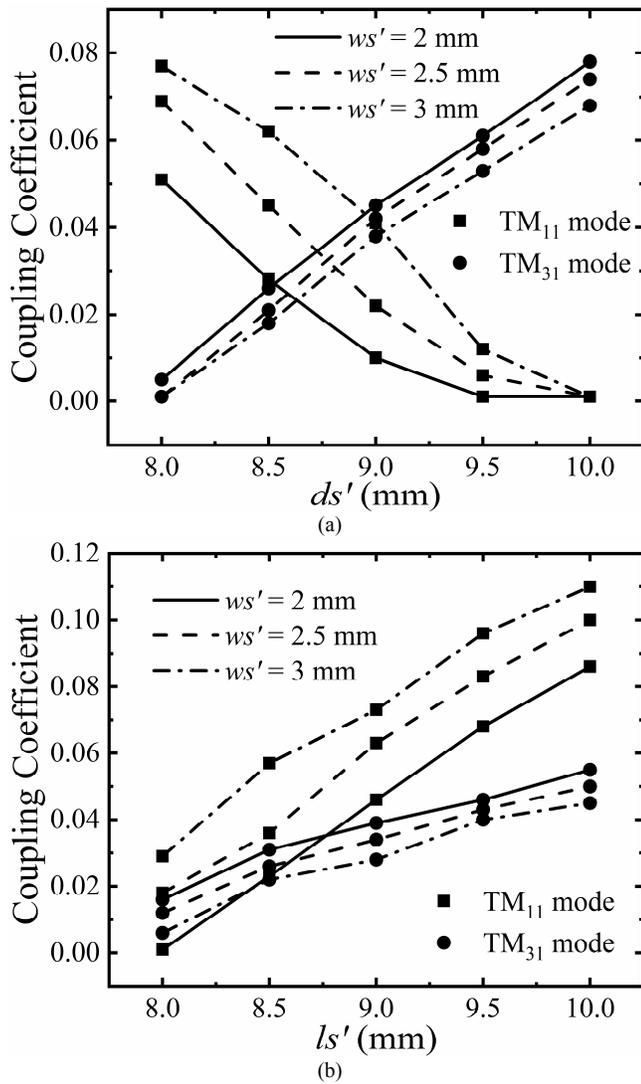

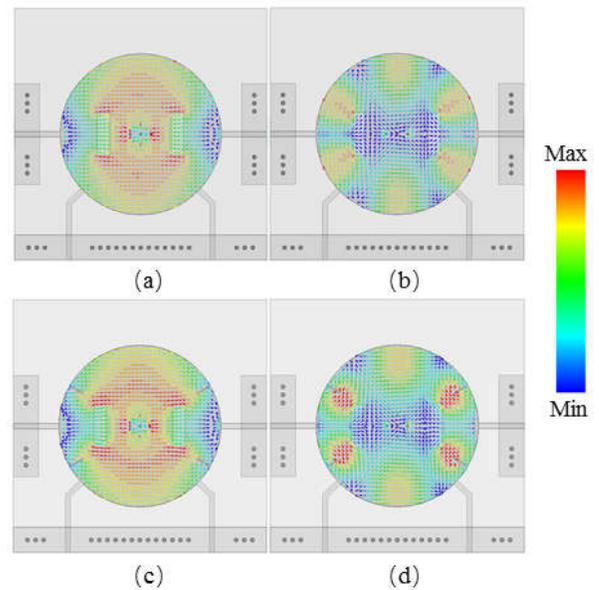

Fig. 13. The coupling coefficients of the proposed dual-band balanced BPF by varying the aperture parameters: (a) $ds'$ ($ls'$ = 8.5 mm); (b) $ls'$ ($ds'$ = 8.6 mm) with different $ws'$ for both $TM_{11}$ and $TM_{31}$ modes.

Fig. 14. The current distributions on the circular patch under the DM excitation: (a) $TM_{11}$ mode without slots at 2.855 GHz; (b) $TM_{31}$ mode without slots at 6.33 GHz; (c) $TM_{11}$ mode with slots ($lc$ = 3 mm) at 2.819 GHz; (d) $TM_{31}$ mode with slots ($lc$ = 3 mm) at 6.015 GHz.

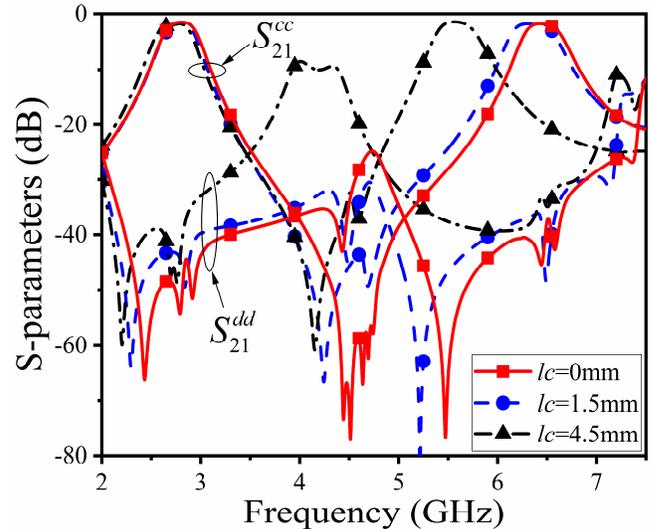

Fig. 15. The adjustability of the second passband of the proposed dual-band balanced BPF by varying the length of the slot cut on the patch ($lc$).

Once again, the couplings of $TM_{11}$ mode and $TM_{31}$ mode between the two circular patch resonators are realized by the two apertures etched on the ground plane as shown in Fig. 9(c). The couplings coefficients calculated from (8) are depicted in Fig. 13 by varying the parameters of the two apertures. It can be seen from Fig. 13(a) that as $ds'$ of the apertures is increased, the couplings of the $TM_{11}$ mode between the resonators is weakened while that of the $TM_{31}$ mode enhanced. While it can be seen from Fig. 13(b) that the couplings of $TM_{11}$ mode and $TM_{31}$ mode under the DM excitation are enhanced together while $ls'$ is increased. The $ws'$ of the apertures has the same effects as $ds'$ upon the couplings of the modes in the circular patch resonators under the DM excitation. These properties can be used to adjust the bandwidth of the DM passbands.

C. *Passbands position adjustment of the dual-band balanced BPF*

As can be seen in Fig. 9(a) and 9(c), there are four slots cuts in each circular patch resonators, which can be used to modify the resonance of the second DM passband. According to Fig. 5(d), the slots are located at where the electric field distributions of $TM_{31}$ mode are zero under the DM excitation. However, in view of (7), the magnetic fields are not zero at these locations. The current distribution of $TM_{11}$ mode and $TM_{31}$ modes under the DM excitation on the circular are plotted in Figs 14(a) and 14(b), respectively. When the slots are perpendicular to the current paths, the current will be interfered as shown in Fig. 14 (c) and 14(d). As a result, the current paths become longer and the resonant frequencies of $TM_{11}$ and $TM_{31}$ modes under the DM excitation decrease. Besides, the current distribution of $TM_{31}$ mode is much stronger than that of $TM_{11}$ mode at the locations of the slots. Therefore, in our design, $\varphi_c$ is chosen to be about $60°$. Fig. 15 depicts the simulated results of the dual-band balanced BPF by varying the length of the slots. Therefore, the first DM passband is kept almost unchanged while the second DM passband can be modified greatly. Although the center frequency of the second DM passband is decreased from 6.45 GHz to 5.54 GHz by increasing $ls$ from 0 to 4.5 mm, the CM suppressions at the two DM passbands remain beyond 35 dB.

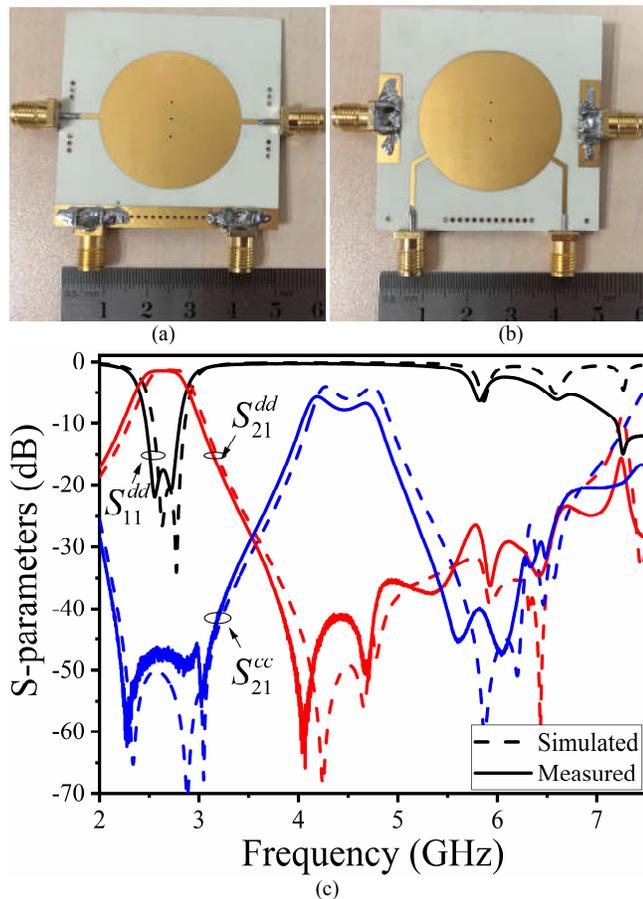

Fig. 16. The (a) top view and (b) bottom view of the fabricated single-band balanced BPF; (c) simulated and measured S-parameters of the fabricated single-band balanced BPF.

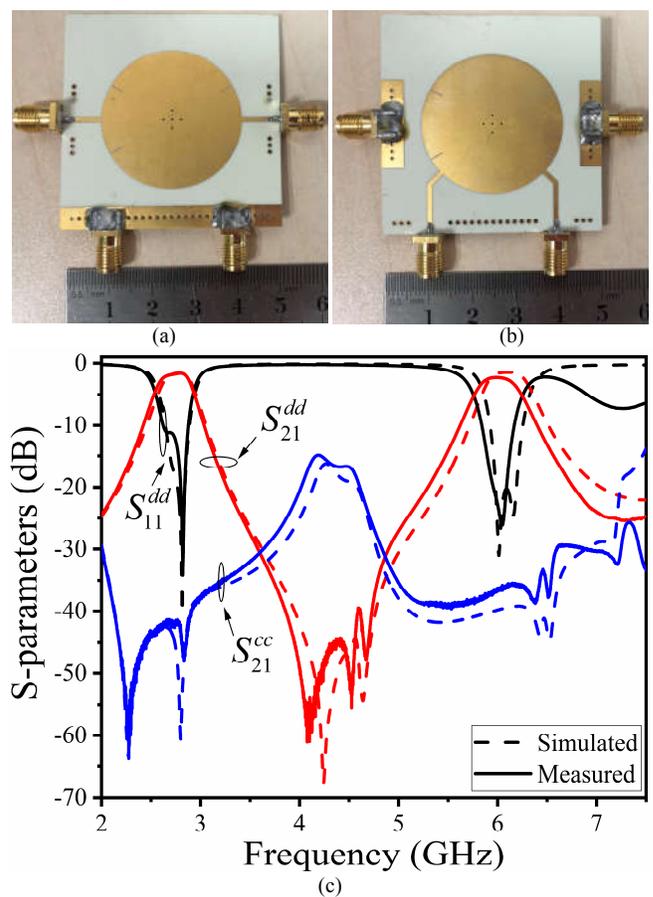

Fig. 17. The (a) top view and (b) bottom view of the fabricated dual-band balanced BPF; (c) simulated and measured S-parameters of the fabricated dual-band balanced BPF.

## IV. EXPERIMENTS AND RESULTS

In this section, the proposed single-band and dual-band balanced BPFs are fabricated and characterized. Figs. 16(a) and 16(b) depict the top and bottom views of the fabricated single-band balanced BPF, respectively. The detailed parameters of the single-band balanced BPF are: $R$ = 16 mm, $\varphi_1 = 60°$, $ws$ = 1 mm, $ls$ = 11.6 mm, $ds$ = 8.7 mm, $s$ = 4 mm, $d$ = 0.4 mm, and $wp$ = 1.05 mm. The simulated and measured results of the single-band balanced BPF are plotted in Fig. 16 (c), which demonstrate very good agreement. For the DM signals, the fabricated single-band balanced BPF operates at the center frequency of 2.63 GHz with a 3 dB fraction bandwidth (FBW) of 15.7% and a minimal in-band return loss of 17.5 dB. The insertion loss is less than 1.5 dB. Besides, there are two TZs located at 4.067 GHz and 4.683 GHz. While for the CM signals, the suppression is beyond 46 dB in the DM passband.

Moreover, Fig. 17(a) and 17(b) depict the top and bottom views of the fabricated dual-band balanced BPF, respectively. Most of the parameters of the dual-band balanced BPF are the same as the single-band one except that: $\varphi_2 = 45°$, $\varphi_c = 60°$, $lc$ = 3 mm, $wc$ = 0.2 mm, $ls'$ = 8.6 mm, $ds'$ = 9 mm, $ws'$ = 2.5 mm, and $s'$ = 2 mm. The simulated and measured results of the dual-band balanced BPF are plotted in Fig. 17(c), which demonstrate very good agreement again. For the DM signals, the dual-band balanced BPF operates at the center frequencies of 2.74 GHz and 6.0 GHz with 3 dB FBWs of 11.0% and 4.8%, respectively. The minimal in-band return losses of the first and second DM passbands are 11 dB and 20 dB, respectively, while the insertion losses of the first and second DM passbands are 1.6 dB and 2.3 dB, respectively. Besides, there are two TZs located at 4.1 GHz and 4.562 GHz. For the CM signals, the suppression is beyond 46 dB in the first DM passband and 35.5 dB in the second DM passband. The slight discrepancies between measurements and simulations are probably due to the fabrication tolerances and the SMA connectors.

In order to demonstrate the desired features of the proposed

TABLE I
COMPARISON WITH OTHER SINGLE-BAND BALANCED BPFS

| Single-band Balanced BPFs | Type | Passband order | Center frequency $f_0^d$ (GHz) | 3-dB FBW(%) | Insertion loss $\left|S_{21}^{dd}\right|$, dB | In-band minimum CM suppression $\left|S_{21}^{dd}\right|$, dB | Effective circuit size $(\lambda_g \times \lambda_g)$ |
|---|---|---|---|---|---|---|---|
| Ref.[5] | Microstrip | 2 | 2.5 | 10 | 1.57 | 34 | 0.15×0.21 |
| Ref.[13]I | Microstrip | 4 | 3.04 | 21.8 | 0.6 | 35 | 9.41×6.40 |
| Ref.[15] | Microstrip | 6 | 3.1 | 21.9 | 1.65 | 22 | 1.05×1.21 |
| Ref.[20] | Patch | 2 | 2.35 | 6 | 1.50 | 50 | 0.78×1.14 |
| Ref.[21] | Patch | 2 | 1.8 | 7 | 1.30 | 30 | 0.55×0.27 |
| Ref.[23] | Patch | 2 | 2.44 | 6.15 | 1.00 | 24 | 1.48×1.48 |
| Ref.[24]I | Patch | 2 | 2.18 | 24.8 | 0.35 | 25 | 0.35×0.70 |
| Ref.[24]II | Patch | 3 | 2.21 | 13.6 | 0.45 | 49.9 | 0.71×0.71 |
| Ref.[24]III | Patch | 5 | 2.22 | 16.6 | 0.55 | 57.7 | 0.71×1.40 |
| **This work** | **Patch** | **2** | **2.63** | **15.7** | **1.50** | **46** | **0.59×0.59** |

DM: differential-mode; CM: common-mode, $\lambda_g$: the guided wave length at the center frequency of the DM passband.

TABLE II
COMPARISON WITH OTHER DUAL-BAND BALANCED BPFS

| Dual-band Balanced BPFs | Type | Passband order | Center frequency $f_{01}^d/f_{02}^d$ (GHz) | 3-dB FBW(%) | Insertion loss $\left|S_{21}^{dd}\right|$, dB | In-band minimum CM suppression $\left|S_{21}^{dd}\right|$, dB | Passband adjustable | Effective circuit size $(\lambda_g \times \lambda_g)$ |
|---|---|---|---|---|---|---|---|---|
| Ref.[6] | Microstrip | 4 | 2.44/5.57 | 16/8.6 | 1.78/2.53 | 35/27 | No | 1.02×2.34 |
| Ref.[8] | Microstrip | 2 | 1.8/5.8 | 10/5.3 | 1.2/2.0 | 35/26 | Yes | 0.37×0.28 |
| Ref.[10] | Microstrip | 4 | 2.5/5.6 | 8/5 | 1.29/1.97 | 40/27 | No | 0.153×0.268 |
| Ref.[11] | Microstrip | 2 | 0.9/2.49 | 3.6/2.1 | 2.67/4.65 | 30/40 | No | 0.67×0.32 |
| Ref.[12] | Microstrip | 2 | 2.4/3.57 | 8/4 | 0.87/1.8 | 28/31 | No | 0.50×0.20 |
| Ref.[13]II | Microstrip | 4 | 2.82/3.21 | 5.2/5.1 | 1.9/1.7 | 15.1/9.3 | No | 11.0×5.19 |
| Ref.[16] | SIW | 4 | 9.23/14.05 | 2.8/5.6 | 2.9/2.7 | 45/40 | Yes | 2.7×1.27 |
| Ref.[17] | SIW | 2 | 6.22/8.24 | 7.7/3.9 | 0.86/1.32 | 41/36 | No | 0.89×0.41 |
| Ref.[18] | SIW | 2 | 3.5/5.24 | 3.1/3.8 | 1.52/1.65 | 55/50 | No | 1.23×1.23 |
| **This work** | **Patch** | **2** | **2.74/6.0** | **11.0/4.8** | **1.6/2.3** | **46/35.5** | **Yes** | **0.59×0.59** |

DM: differential-mode; CM: common-mode, $\lambda_g$: the guided wave length at the center frequency of the first DM passband.

balanced BPFs, the comparisons of various performances among the proposed single-band (dual-band) balanced BPF and other balanced BPFs reported in the literature are summarized in Table I (Table II). It can be concluded from Table I that the proposed single-band balanced BPF features a compact size and good CM suppression compared to other second-order single-band balanced BPFs. Moreover, the proposed dual-band balanced BPF can be achieved by slightly modifying the proposed single-band balanced BPF, which also exhibits good features in comparison with other dual-band balanced BPFs, such as compact size, high CM suppression, and adjustable passbands. In addition, by suppressing $TM_{21}$ mode under the DM excitation, two TZs (4.067 GHz and 4.683 GHz in the single-band balanced BPF, 4.1 GHz and 4.562 GHz in the dual-band balanced BPF) around the resonant frequency of $TM_{21}$ mode could be introduced to improve the out-band-performance of the single- and dual-band balanced BPFs under the DM operation.

## V. CONCLUSION

In this work, the novel single-band and dual-band balanced BPFs are proposed based on the circular patch resonators. The characteristics of the resonant modes of the circular patch resonator are investigated and analyzed. By modifying the output balanced ports, a single-band or dual-band balanced BPF with or without suppressing the $TM_{31}$ mode can be achieved. Meanwhile, the even modes ($TM_{21}$ and $TM_{01}$ modes) restrained under the DM excitation could produce two TZs at the upper sideband of the of the single-band balanced BPF or between the two DM passbands of the dual-band balanced BPF, which could lead to good out-of-band rejections. In addition, the CM suppression is kept beyond a high level in the DM passbands due to the odd modes ($TM_{11}$ and $TM_{31}$ modes) are suppressed under the CM excitation. Moreover, perturbation vias are introduced to disturb the electric field distributions of the even modes under the CM excitation to improve CM suppression around the DM passbands, and slots are cut on the circular patches to perturb the current distribution of the odd modes under the DM excitation to adjust the second DM passband of the dual-band balanced BPF. Finally, the proposed single-band and dual-band balanced BPFs are fabricated and measured. The measured results agree very well with the simulated ones, which validates the proposed method. The proposed designs feature compact size, high CM suppression and adjustable passbands, which could have great potential in the wireless communication and radar systems.